# Warranty Cost Analysis with an Alternating Geometric Process

Richard Arnold[1], Stefanka Chukova[1], Yu Hayakawa[2], Sarah Marshall[3*]

1. School of Mathematics and Statistics,
Victoria University of Wellington PO Box 600,
Wellington 6140, New Zealand
2. School of International Liberal Studies,
Waseda University, 1-6-1 Nishi-Waseda,
Sinjuku-ku, Tokyo 169-8050, Japan
3. School of Engineering, Computer and Mathematical Sciences,
Auckland University of Technology
Private Bag 92006, Auckland 1142, New Zealand

**ABSTRACT**

In this study we model the warranty claims process and evaluate the warranty servicing costs under non-renewing and renewing free repair warranties. We assume that the repair time for rectifying the claims is non-zero and the repair cost is a function of the length of the repair time. To accommodate the ageing of the product and repair equipment, we use a decreasing geometric process to model the consecutive operational times and an increasing geometric process to model the consecutive repair times. We identify and study the alternating geometric process (AGP), which is an alternating process with cycles consisting of the item's operational time followed by the corresponding repair time. We derive new results for the AGP in finite horizon and use them to evaluate the warranty costs over the warranty period and over the life cycle of the product under a non-renewing free repair warranty (NRFRW), a renewing free repair warranty (RFRW) and a restricted renewing free repair warranty (RRFRW($n$)). Properties of the model are demonstrated using a simulation study.

**KEYWORDS:** warranty cost analysis; geometric process; alternating geometric process

**NOMENCLATURE**

$\{X_i\}_1^\infty$ – a stochastically decreasing geometric process with parameters $\{a, F_{X_1}(t)\}$, $a \geq 1$, representing the "on" times;

$\{Y_i\}_1^\infty$ – a stochastically increasing geometric process with parameters $\{b, F_{Y_1}(t)\}$, $0 < b \leq 1$, representing the "off" times;

$Z_i$ – the length of the $i^{th}$ cycle, with the cumulative distribution function $H_i(t)$;

$S_n$ – the end of the $n^{th}$ cycle, with the cumulative distribution function $G_i^n(t)$;

$T$ – the length of the warranty period;

$L$ – the length of the life cycle;

$C(T)$ – the warranty cost over the warranty period;

$C(L)$ – the warranty cost over the life cycle.

## 1. INTRODUCTION

In warranty cost analysis it is typically assumed that the time required to rectify a warranty claim is negligible. In many cases this assumption is reasonable but there are situations where this assumption is hard to justify, e.g., having lengthy repairs with high penalty costs or lengthy repair leading to substantial loss of income. In these cases, ignoring the length of repair will lead to underestimation of the expected warranty costs.

Chukova and Hayakawa [1, 2] studied models based on an alternating renewal process (i.e., assuming independent and identically distributed (i.i.d.) operational times and independent and identically distributed repair times) to evaluate the warranty costs under both non-renewing and renewing warranties. Non-zero repair times and a finite time horizon were taken into account in these models. For details on alternating renewal processes, see

* Corresponding Author: sarah.marshall@aut.ac.nz
Tel: +64-9-921-9999-5414.

[3]. For more on the application of renewal theory in warranty cost analysis, see [4]. For more on models with non-zero repair times see [5]–[12].

The main goal of this paper is to study a generalization of the aforementioned results in [1, 2] to more realistically account for product ageing, i.e., the operational times are stochastically decreasing and the repair times are stochastically increasing. We use the geometric process to model the stochastically increasing and decreasing times (see [13] for an overview of the geometric process). In this paper we model a scenario in which, as an item ages, the operational time decreases and the time required to bring a faulty product to a functioning condition increases. Hence, the new model will be based on an alternating geometric process (AGP), which we will introduce in Section 2.

## 2. THE MODEL

### 2.1. Alternating geometric process (AGP)

Consider an item, which initially operates for a length of time $X_1$ and then fails. After this, it undergoes repair for a length of time $Y_1$. After the repair, the item is again operational for a time $X_2$, which is followed by a repair for a time $Y_2$ and so on. We assume that:

(1) $\{X_i\}_1^\infty$ and $\{Y_i\}_1^\infty$ are independent sequences of random variables;
(2) $\{X_i\}_1^\infty$ is a stochastically decreasing geometric process with parameters $\{a, F_{X_1}(t)\}$, $a \geq 1$;
(3) $\{Y_i\}_1^\infty$ is a stochastically increasing geometric process with parameters $\{b, F_{Y_1}(t)\}$, $0 < b \leq 1$.

A stochastic process $\{Z_i\}_1^\infty$ is referred to as a geometric process with parameter $\beta$ if there exists a real number $\beta > 0$ such that $\{\beta^{i-1} Z_i\}_1^\infty$ is a renewal process [13]. A geometric process is stochastically increasing if $0 < \beta \leq 1$ and stochastically decreasing if $\beta \geq 1$. If $\beta = 1$, then the process becomes a renewal process. See [14] for a different parametrization of the geometric process. The process described above is referred to as an alternating geometric process (AGP) with parameters $\{a, F_{X_1}(t), b, F_{Y_1}(t)\}$.

We refer to a period of time as a "cycle" if it consists of an operational ("on") time followed by the corresponding repair ("off") time. We suppose that the repair cost is incurred at the end of each cycle. If the warranty coverage expires during a repair period, the corresponding repair is completed and its cost is fully incurred by the warrantor. In this case we have a complete cycle. If the warranty expires during an operational period, the cost of the following repair is not covered by the warrantor and the cycle is incomplete.

Similar to [1, 2], we assume that the cost of the $i^{th}$ repair has the form $C_i = A + \delta Y_i$, where $A$ and $\delta$ are prespecified constants.

The life cycle of a product is defined as the time while the product is still usable and contemporary. We assume that during the life cycle, after the expiration of the warranty coverage for the initially purchased item, at the time of the first off-warranty failure, the consumer purchases an identical item to the initial one, with the same warranty coverage.

### 2.2. AGP in finite horizon

Consider an AGP with the $i^{th}$ "on" time distribution $F_{X_i}$ and $i^{th}$ "off" time distribution $F_{Y_i}$. We assume that the "on" and "off" time processes are geometric processes. The "on" time process is a decreasing geometric process with parameters $a \geq 1$, $F_{X_i}(t) = F_{X_1}(a^{i-1}t), i = 1, 2, ...$. The "off" time process is an increasing geometric process with parameters $0 < b \leq 1$, $F_{Y_i}(t) = F_{Y_1}(b^{i-1}t), i = 1, 2, ...$. Denote by $Z_i = X_i + Y_i$, the length of the $i^{th}$ cycle, i.e., the sum of the $i^{th}$ operational and $i^{th}$ repair times, with the cumulative distribution function $H_i(t)$. Let $S_n = \sum_{i=1}^n (X_i + Y_i)$. Then, the number of AGP cycles completed by time $t$, $N(t)$, and its expected value, $m_1(t)$, are given respectively by

$$N(t) = \sup\{n: S_n \leq t\} \text{ and } m_1(t) = E(N(t)).$$

Analogously to computing the renewal function (see [3]), we can see that $m_1(t)$ can be represented as

$$m_1(t) = \sum_{n=1}^\infty P(S_n \leq t) = \sum_{n=1}^\infty G_1^n(t),$$

where

$$G_i^n(t) = H_i * H_{i+1} * \ldots * H_{i+n-1} \tag{1}$$

and "$*$" denotes a convolution.

Next, we summarize some of the results needed to evaluate the expected warranty costs. Most of the results are stated without proof. Firstly, by extending the results of [13][Thm 2.3.1], the probability that the system is "on" at time $t$ can be obtained as follows

$$P(\text{on at } t) = \bar{F}_{X_1}(t) + \sum_{n=1}^\infty \int_0^t \bar{F}_{X_{n+1}}(t-s) \, dG_1^n(s). \tag{2}$$

Let $T > 0$ be the length of a finite period of time. Then, the following results hold:

**Theorem 1.**
$$E(Y_{N(T)+1} | \text{on at } T)$$
$$= \frac{E(Y_1)}{P(\text{on at } T)} \left\{ \bar{F}_{X_1}(t) + \sum_{n=1}^\infty \frac{1}{b^n} \int_0^T \bar{F}_{X_{n+1}}(T-s) \, dG_1^n(s) \right\}. \tag{3}$$

**Theorem 2.**

$$P(S_{N(T)} + X_{N(T)+1} \leq t \mid \text{on at } T)$$
$$= \frac{\bar{F}_{X_1}(T) - \bar{F}_{X_1}(t)}{\bar{F}_{X_1}(T) + \sum_{n=1}^{\infty} \int_0^T \bar{F}_{X_{n+1}}(T-s) \, dG_1^n(s)}$$
$$+ \frac{\sum_{n=1}^{\infty} \int_0^T \left( \bar{F}_{X_{n+1}}(T-s) - \bar{F}_{X_{n+1}}(t-s) \right) dG_1^n(s)}{\bar{F}_{X_1}(T) + \sum_{n=1}^{\infty} \int_0^T \bar{F}_{X_{n+1}}(T-s) \, dG_1^n(s)}$$
(4)

**Theorem 3.**

$$P(S_{N(T)+1} + X_{N(T)+2} \leq t \mid \text{off at } T) = \frac{1}{P(\text{off at } T)}$$
$$\times \left( \int_0^T \int_{T-u}^{t-u} F_{X_2}(t-u-v) \, dF_{Y_1}(v) \, dF_{X_1}(u) \right.$$
$$+ \sum_{n=1}^{\infty} \int_0^T \int_0^{T-s} \int_{T-s-u}^{t-s-u} F_{X_{n+2}}(t-s-u$$
$$\left. - v) \, dF_{Y_{n+1}}(v) \, dF_{X_{n+1}}(u) \, dG_1^n(s) \right)$$
(5)

### 3. WARRANTY COST ANALYSIS UNDER AN NRFRW

Next, we consider a non-renewing free repair warranty (NRFRW), i.e., the product is warrantied for a fixed period of time $T$, usually starting right after the purchase. During the warranty period all expenses are borne by the manufacturer. We derive the expected warranty costs over the warranty period $T$ and over the life cycle of length $L$.

#### 3.1. Expected warranty costs over (0, T)

The total cost over the warranty period, $C(T)$, can be represented as

$$C(T) = \begin{cases} \sum_{i=1}^{N(T)} C_i & \text{if "on" at time } T \\ \sum_{i=1}^{N(T)+1} C_i & \text{if "off" at time } T. \end{cases}$$

Then,

$$E(C(T)) = E\left( \sum_{i=1}^{N(T)+1} C_i \right) - E(C_{N(T)+1} \mid \text{on at } T) P(\text{on at } T),$$
(6)

where $P(\text{on at } T)$ is given by (2). For $b \neq 1$, we have

$$E(C_i) = E(A + \delta Y_i) = A + \delta \frac{E(Y_1)}{b^{i-1}},$$
(7)

$$E(C_{N(T)+1} \mid \text{on at } T) = A + \delta E(Y_{N(T)+1} \mid \text{on at } T),$$
(8)

and,

$$E\left( \sum_{i=1}^{N(T)+1} C_i \right) = E\left( \sum_{i=1}^{N(T)+1} (A + \delta Y_i) \right)$$
$$= A(m_1(T) + 1) + \delta E\left( \sum_{i=1}^{N(T)+1} Y_i \right)$$
$$= A(m_1(T) + 1) + \delta \frac{E(Y_1)\{E(b^{-N(T)}) - b\}}{1 - b}.$$
(9)

Note that when $b = 1$, the repair time process is a renewal process. For the expected warranty cost under this scenario refer to [1].

#### 3.2. Expected warranty costs over (0, L)

Let $L^*$ be a prespecified time during which a product is considered to be contemporary and competitive with similar products in the market. Let $L$ be the time of the first off-warranty failure of the product after $L^*$. Then, we call $(0, L)$ the life-cycle of the product. Let $\xi$ be a positive random variable, representing the time between two consecutive purchases, i.e.,

$$\xi = \begin{cases} S_{N(T)} + X_{N(T)+1}, & \text{if "on" at time } T \\ S_{N(T)+1} + X_{N(T)+2}, & \text{if "off" at time } T \end{cases}$$

Then, the expected cost over $(0, L)$ is given by

$$E(C(L)) = (m_\xi^*(L) + 1) E(C(T)),$$

where $m_\xi^*(t)$ is the renewal function of the renewal process generated by $\xi$. Based on the definition of $\xi$, its distribution can be represented via the respective conditional distributions of $S_{N(T)} + X_{N(T)+1}$ and $S_{N(T)+1} + X_{N(T)+2}$, given in (4) and (5) respectively.

### 4. WARRANTY COST ANALYSIS UNDER A RFRW AND A RRFRW(n)

Next, we consider a renewing free repair warranty (RFRW) under which, following a warranty repair, the item is warranted anew for a period of length $T$. If the warranty period ends during an operational period, the cost of the following repair is not incurred by the warrantor and the warranty coverage expires. Here we will distinguish between the warranty coverage $W_T$, which is a random variable, and the warranty period, which is a predetermined constant $T$. We define $W_T$ as the time from the purchase of the product until the expiry of the warranty coverage. We also consider a restricted renewing free repair warranty (RRFRW($n$)), in which the number of warranty repairs is limited to some predetermined number $n$. We define $W_T^n$ as the warranty coverage under a RRFRW($n$).

### 4.1. Cost analysis for a RFRW

#### 4.1.1. Expected warranty costs over $(0, W_T)$

Due to the mechanism of the renewing warranty, $W_T$ is equal to:

$$W_T = \begin{cases} T & \text{if } X_1 > T \\ T + \sum_{i=1}^{k}(X_i + Y_i) & \text{if } X_i \leq T, i = 1, 2, \dots, k, \\ & X_{k+1} > T, \text{ for some } k. \end{cases}$$
(10)

Then, the warranty cost $C(W_T)$ over the warranty coverage is a random variable and its distribution is as follows:

$$C(W_T) = \begin{cases} 0, & \text{with probability } 1 - F_{X_1}(T) \\ C_1 & \text{with probability } \left(1 - F_{X_2}(T)\right) F_{X_1}(T) \\ C_1 + C_2 & \text{with probability } \left(1 - F_{X_3}(T)\right) F_{X_2}(T) F_{X_1}(T) \\ \vdots & \vdots \\ \sum_{i=1}^{k} C_i & \text{with probability } (1 - F_{X_{k+1}}(T)) \prod_{i=1}^{k} F_{X_i}(T) \\ \vdots & \vdots \end{cases}$$
(11)

where $E(C_i)$, for $b \neq 1$, is given in (7).

Next, we consider the expected warranty costs $E(C(W_T))$ over $(0, W_T)$. After some algebraic manipulations, we obtain that

$$E(C(W_T)) = \sum_{k=1}^{\infty} \left( A + \frac{\delta E(Y_1)}{b^{k-1}} \right) \prod_{j=1}^{k} F_{X_j}(T).$$
(12)

It can be shown that the series (12) is divergent (based on d'Alembert's test and the Stolz-Cesàro theorem [15]). Hence, $E(C(W_T))$ goes to infinity. Therefore, assigning a warranty period of length $T$ for a product, with operational/repair times that form an AGP with parameters $\{a, F_{X_1}(t), b, F_{Y_1}(t)\}$ is not a viable business option. It may, however, be practical to offer a restricted renewing free repair warranty, with at most $n$ warranty repairs (RRFRW($n$)), which we will present in subsection 4.2.

### 4.2. Cost analysis for a RRFRW($n$)

As an alternative to a RFRW strategy we will consider its modified version, called a restricted renewing free repair warranty with parameter $n$ (RRFRW($n$)), under which at most $n$ warranty repairs are covered by the warranty coverage, where $n$ is a known, fixed constant.

#### 4.2.1. Expected warranty costs over $(0, W_T^n)$

Under a RRFRW($n$), the warranty coverage $W_T^n$ can be represented as follows:

$$W_T^n = \begin{cases} T & \text{if } X_1 > T \\ \sum_{i=1}^{k-1}(X_i + Y_i) + T & \text{if } X_i \leq T, i = 1, 2, \dots, k-1, \\ & X_k > T, 2 \leq k \leq n \\ \sum_{i=1}^{n}(X_i + Y_i) & \text{if } X_i \leq T, i = 1, 2, \dots, n. \end{cases}$$

Then, the warranty cost $C(W_T^n)$ over the warranty coverage is a random variable and its distribution is as follows:

$$C(W_T^n) = \begin{cases} 0, & \text{with probability } 1 - F_{X_1}(T) \\ C_1 & \text{with probability } \left(1 - F_{X_2}(T)\right) F_{X_1} \\ \vdots & \vdots \\ \sum_{i=1}^{k-1} C_i & \text{with probability } \left(1 - F_{X_k}(T)\right) \prod_{i=1}^{k-1} F_{X_i}(T), \\ & \text{for } k < n \\ \vdots & \vdots \\ \sum_{i=1}^{n-1} C_i & \text{with probability } \left(1 - F_{X_n}(T)\right) \prod_{i=1}^{n-1} F_{X_i}(T) \\ \sum_{i=1}^{n} C_i & \text{with probability } \prod_{i=1}^{n} F_{X_i}(T). \end{cases}$$

It is easy to derive that under a RRFRW($n$) the expected warranty cost is given by

$$E(C(W_T^n)) = \sum_{k=1}^{n} \left( A + \frac{\delta E(Y_1)}{b^{k-1}} \right) \prod_{j=1}^{k} F_{X_j}(T),$$
(13)

which is a truncated version of the divergent series (12) to its $n^{\text{th}}$ partial sum. Hence, for a RRFRW($n$) under an AGP with parameters $\{a, F_{X_1}(t), b, F_{Y_1}(t)\}$, the expected warranty cost is always finite, and this warranty strategy might be considered as an appropriate warranty strategy by some producers.

### 4.2.2. Expected warranty costs over $(0, L)$

Similar to section 3.2, expressions for the expected warranty costs over $(0, L)$ can be derived under a RRFRW($n$). Consider the positive random variable $\xi^n$ as the time between two consecutive purchases under a RRFRW($n$). By definition

$$\xi^n = \begin{cases} X_1 & \text{if } X_1 > T \\ \sum_{i=1}^{k-1}(X_i + Y_i) + X_k & \begin{array}{l}\text{if } X_i \leq T, i = 1,2,\ldots,k-1, \\ X_k > T, \ 2 \leq k \leq n\end{array} \\ \sum_{i=1}^{n}(X_i + Y_i) + X_{n+1} & \text{if } X_i \leq T, i = 1,2,\ldots,n. \end{cases}$$

Then, it can be shown that the cdf of $\xi^n$ is given by

$$P(\xi^n \leq t) = P(T < X_1 \leq t) + \\ \int_0^{t-T} P(T < X_2 \leq t-s) \ dG_1^1(s) + \\ \int_0^{t-T} P(T < X_3 \leq t-s) \ dG_1^2(s) + \cdots \\ \int_0^{t-T} P(T < X_n \leq t-s) \ dG_1^{n-1}(s) + \\ \int_0^{t} P(X_{n+1} \leq t-s) \ dG_1^n(s).$$

Therefore, the following theorem holds:

**Theorem 5.**

$$F_{\xi^n}(t) = (F_{X_1}(t) - F_{X_1}(T)) + \\ \sum_{i=1}^{n-1} \int_0^{t-T} \left(F_{X_{i+1}}(t-s) - F_{X_{i+1}}(T)\right) dG_1^i(s) + \\ \int_0^{t} \left(F_{X_{n+1}}(t-s)\right) dG_1^n(s),$$

where $G_1^i(s)$ is given by (1).

Then, the expected warranty costs over $(0, L)$, say $E(C(L))$, are expressed in terms of $\xi^n$ in the following way

$$E\big(C(L)\big) = \big(m_{\xi^n}^*(L) + 1\big) E\big(C(W_T^n)\big),$$

where $m_{\xi^n}^*$ is the renewal function of the renewal process generated $\xi^n$.

## 5. SIMULATION RESULTS FOR A NRFRW

In this section, using simulation, we study the expected warranty cost for a NRFRW under an AGP with parameters $\{a, F_{X_1}(t), b, F_{Y_1}(t)\}$. The expected warranty costs over the warranty period, as well as the life cycle, are estimated using the average cost over 5 million warranty cost simulations under a NRFRW.

The expected warranty cost $E(C(T))$, given in (6), is influenced mainly by the following two factors:
 i. the number of claims and
 ii. the cost of these claims.

The cost of the claims depends on the length of the repair time (driven by the repair time distribution $F_{Y_1}$ and $0 < b < 1$). The number of claims depends on the operational time (driven by operational time distribution $F_{X_1}$ and $a > 1$) as well as the repair times. Recall that the cost of the $i^{\text{th}}$ claim is $C_i = A + \delta Y_i$, where $A$ could be interpreted as a fixed cost incurred for each repair and $\delta$ could be interpreted as a variable cost per time unit.

Figure 1 explores the relationship between the repair rate $\mu = 1/E(Y_1)$ and warranty cost for an exponential repair time distribution for various values of the cost parameter $\delta$. The operational times are modelled by an exponential distribution with rate $\lambda = 1/E(X_1) = 0.0055$. If the time unit for the simulation is a day, then this corresponds to a failure rate of approximately 2 per year and thus an average operational time of about 182 days. A repair rate of $\mu = 1/E(Y_1) = 2$ corresponds to an average repair time of 0.5 days and a repair rate of $\mu = 1/E(Y_1) = 0.01$ corresponds an average repair time of 100 days. As expected, a comparison of the three graphs in Figure 1 shows that as the cost parameter $\delta$ increases, the expected warranty cost will increase. Notice that an increase by a factor of 100 from $\delta = 0.01$ to $\delta = 1$ leads to a similar percentage increase in the warranty costs for small values of $b$. The shape of the graphs, however, changes for different values of $\delta$.

It might be expected that as the repair rate increases (i.e., the length of the repair time decreases) the cost would decrease. That is, we may expect the expected warranty cost for $\mu = 2$ to be less than the cost for $\mu = 0.01$. This can be observed in the third graph of Figure 1 for $\delta = 1$. However, in the first two graphs of Figure 1, for high values of $b$ this is not the case. This can be explained by examining Figure 2. When $b$ is low and $\mu$ is low the expected number of cycles is much lower than when $b$ is high and $\mu$ is high, and correspondingly, the expected cycle length is much larger. In Figure 1, the first graph has a very small variable cost parameter $\delta = 1/365 = 0.00274$ compared with the fixed cost parameter $A = 1$, so the number of cycles dominates the expected warranty cost. Notice that the cost is comparable to the expected number of cycles shown in Figure 2. In the third graph in Figure 1, for $\delta = 1$, the expected warranty cost is dominated by the repair time, rather than the number of cycles, and thus the cost is comparable to the average cycle length shown in Figure 2.

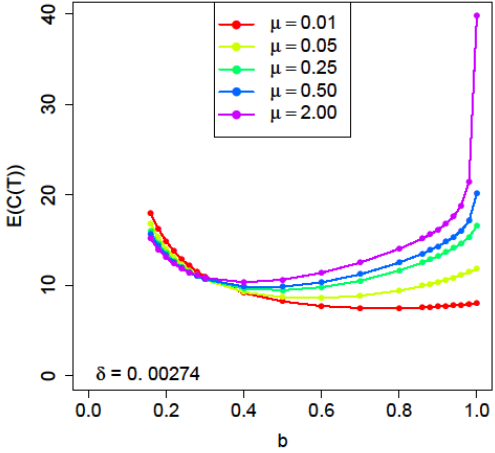

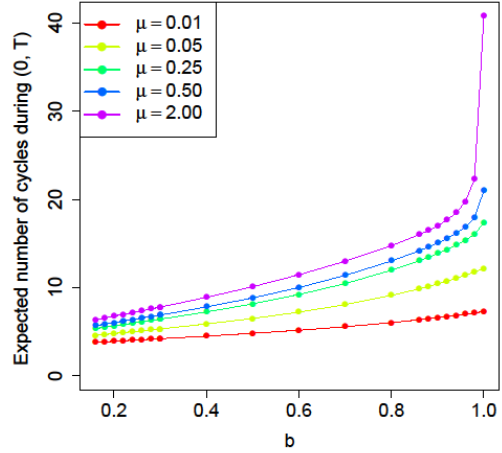

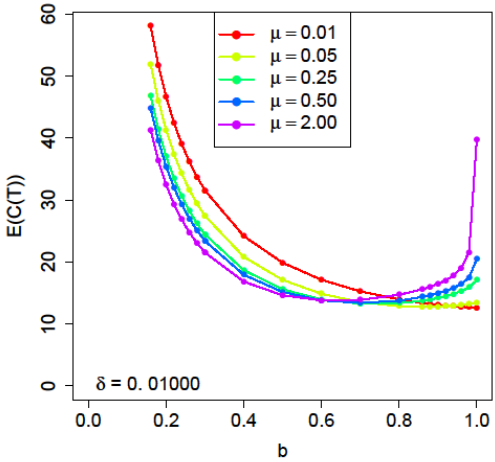

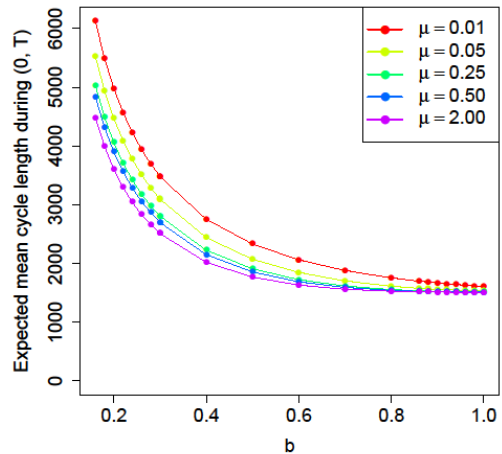

Figure 2. Properties of the cycle with
$X_1$~Exponential($\lambda$=0.0055), $A$=1, $\delta$=1, $T$=1460, $a$=1.1,
$Y_1$~Exponential($\mu$)

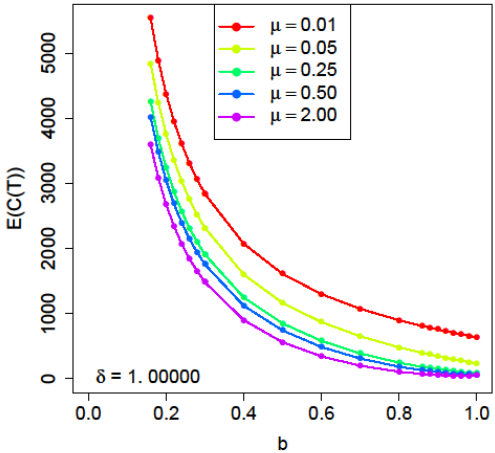

Figure 1. Expected Warranty Cost with
$X_1$~Exponential($\lambda$=0.0055), $A$=1, $T$=1460, $a$=1.1,
$Y_1$~Exponential($\mu$), for various $\delta$

Figure 3 shows that as the parameter *a* increases (i.e., the operational times decrease more rapidly) the expected warranty cost increases. This is to be expected since shorter operational times mean that there will be more claims within the warranty period. The operational times are also influenced by the failure rate $\lambda = 1/E(X_1)$. Figure 3 also shows that higher values of $\lambda$ (i.e., shorter operational times) lead to higher warranty costs. It also shows that for higher values of $\mu$ (i.e., shorter repair times), the expected warranty cost is lower for values of *a*, which are close to 1. For larger values of *a* (approximately $a > 1.3$) and $\lambda$ ($\lambda = 0.0055$ and $\lambda = 0.01$), the expected warranty cost is higher for larger values of $\mu$ (shorter repair times). This suggests that as operational times decrease (larger *a* and $\lambda$) and repair times decrease (larger $\mu$), the cost is driven by the frequency of repairs rather than the length of the repairs.

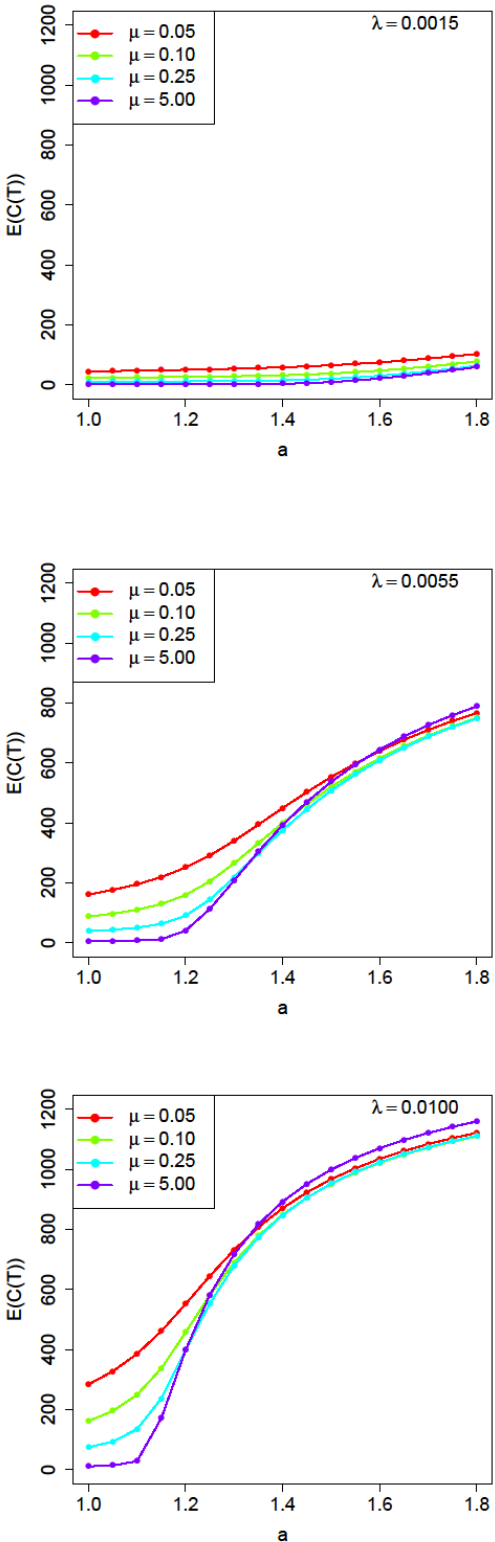

Figure 3 Expected Warranty Cost with $X_1 \sim$ Exponential($\lambda$), $A$=1, $\delta$=2, $T$=730, $b$=0.95, $Y_1 \sim$ Exponential($\mu$), for various $\lambda$

# 6 CONCLUSIONS

In this paper, we studied non-renewing, renewing and restricted renewing free repair warranties under a new failure/repair process based on an alternating geometric process (AGP), which accounts for the shortening behavior of the operational times and for the lengthening behavior of the repair times. Using an AGP, we derived the expected warranty costs over the warranty period and over the life cycle for NRFRW, RFRW and RRFRW($n$) models. Also, using simulation, we demonstrated some properties of the NRFRW model. In our future work, we will explore the properties of the RRFRW($n$) using simulation and work on providing some insight on the statistical inference for the proposed warranty models.

## ACKNOWLEDGMENTS

This research was supported by Waseda University, Grant for Special Research Projects (2016B-267, 2017B-325).